\documentclass[english,twocolumn,american,preprintnumbers,nofootinbib]{revtex4}
\usepackage{ae}
\usepackage{aecompl}
\usepackage[T1]{fontenc}
\usepackage[latin1]{inputenc}
\usepackage{verbatim}
\usepackage{amsmath}
\usepackage{amssymb}

\makeatletter

\makeatletter

\def\be{\begin{equation}}
\def\ee{\end{equation}}
\def\bea{\begin{eqnarray}}
\def\eea{\end{eqnarray}}


\makeatother

\usepackage{babel}
\makeatother
\begin{document}

\preprint{YITP-08-90}

\title{Stationary Measure in the Multiverse}

\author{Andrei Linde$^{1,3}$, Vitaly Vanchurin$^{2,3}$, and Sergei Winitzki$^{2,3}$}

\date{November 29, 2008}

\affiliation{$^{1}$Department of Physics, Stanford University, Stanford, CA 94305,
USA}

\affiliation{$^{2}$Department of Physics, Ludwig-Maximilians University, Munich,
Germany}

\affiliation{$^{3}$Yukawa Institute of Theoretical Physics, Kyoto University,
Kyoto, Japan}

\begin{abstract}
We study the recently proposed {}``stationary measure'' in the context
of the string landscape scenario. We show that it suffers neither from
the {}``Boltzmann brain'' problem nor from the {}``youngness''
paradox that makes some other measures predict a high CMB temperature
at present. We also demonstrate a satisfactory performance of this
measure in predicting the results of local experiments, such as proton
decay. 
\end{abstract}
\maketitle

\section{Introduction}

The great goal of quantum cosmology was to find the wave function
of the universe describing space of all possible geometries, which
was called {}``superspace'' \cite{Wheeler:1964,DeWitt:1967yk}.
Unfortunately, this task proved to be extremely complicated. Many
advances of quantum cosmology are based on investigation of a {}``minisuperspace,''
which describes quantum evolution of a homogeneous and isotropic Friedmann
universe (e.g.~\cite{Vilenkin:1982de,Hartle:1983ai,Linde:1983mx,Vilenkin:1984wp,Linde:2004nz}).

Inflationary cosmology provided a simple explanation of the local
homogeneity of the universe but simultaneously yielded a picture of
the universe consisting of a large number of exponentially large parts
with different properties \cite{Linde:1982ur,Linde:1982gg}. This
picture became even more interesting and unusual with the development
of the theory of eternal inflation~\cite{Vilenkin:1983xq,Linde:1986fd}.
In particular, the first paper on the chaotic eternal inflation ~\cite{Linde:1986fd}
contained the following manifesto:

``From this point of view, an enormously large number of possible types of compactification which exist e.g.~in the theories of superstrings should be considered not as a difficulty but as a virtue of these theories, since it increases the probability of the existence of mini-universes in which life of our type may appear.''  Recently this idea attracted attention of the broad scientific community, when it became a part of the string theory landscape scenario~\cite{Lerche:1986cx,Bousso:2000xa,Kachru:2003aw,Susskind:2003kw,Douglas:2003um}.

According to the new cosmological paradigm, the universe, called more
appropriately a {}``multiverse,'' is an eternally growing fractal
consisting of infinitely many exponentially large parts with different
coupling constants, particle masses, the effective cosmological constant,
and other {}``constants of nature.'' These parts were called {}``mini-universes''
\cite{Linde:1982ur,Linde:1982gg} or {}``pocket universes'' \cite{Guth:2000ka}.

The general structure of the multiverse is extremely complicated;
one cannot describe the growing fractal in terms of a single Friedmann
universe or in the minisuperspace approach. Fortunately, inflation
makes each part of the multiverse locally homogeneous. Rapid expansion
of the universe during inflation makes the evolution of the parts
of the universe of a size much greater than horizon practically independent
of each other. Therefore one can investigate the formation of different
exponentially large parts of the universe using the theory of tunneling
\cite{Coleman:1980aw} and a stochastic approach to inflation \cite{Starobinsky:1986fx},
study each pocket as a locally homogeneous Friedmann universe, and
then investigate different properties of the growing fractal by bringing
together information about all pockets. 
One can call this method a
``multisuperspace'' approach. This approach is closely related to the ideology of the string theory landscape, which emphasizes the possibility of choice between an enormously large number of different vacua of string theory.

 This approach allows further generalizations. First of all, the final outcome of the inflationary evolution depends not only on the properties of the vacuum in the string theory landscape, but also on the inflationary trajectory which brings us there \cite{Aguirre:2006na}. For example, if one of the transitions to the same vacuum state involves tunneling and another involves slow-roll inflation, then perturbations of metric with flat spectrum are produced only in those parts of the universe which are produced by the second mechanism.

More generally, we may be interested not only in predicting  the average properties of a pocket universe, but also in its detailed structure,
geometry and a spatial distribution of the scalar fields. Then we should take into account a detailed history of the evolution of the classical scalar fields and their quantum fluctuations along each inflationary trajectory.

Consider for example a quantum jump which the inflaton field experienced in a vicinity of a given point at the time corresponding to $N$ e-folds preceding the end of inflation. A quantum jump  against the classical motion of the inflaton field  produces a local increase of energy density and the corresponding change in the space-time geometry on scale $\sim H^{-1} e^{N}$ in the neighborhood of this point at the end of inflation. On the other hand, a quantum  jump  along the classical motion of the scalar field produces a local decrease of energy density on the same scale. Note that this modification of the space-time geometry on scale $\sim H^{-1} e^{N}$ cannot be erased by quantum fluctuations which occur at any other stage of inflation: Fluctuations produced much earlier create perturbations on exponentially greater scale, whereas the fluctuations produced later create inhomogeneities on an exponentially smaller scale. 

In other words, each quantum inflationary trajectory produces a distinguishable classical outcome in terms of the distribution of matter and large-scale geometry. This observation, which will play an important role in some of our subsequent considerations,  brings the theory of an inflationary multiverse one step closer to the ultimate goal of the original superspace approach.

Since the fundamental theory admits a wide range of possible values
of the {}``constants of nature'' and geometries, one can only hope to obtain the
\emph{probability} of observing a local part of the universe with a given set of properties. It
is natural to assume that our universe is located at a random place
in the multiverse and that some probability distribution exists for
the {}``typically observed'' values of the cosmological parameters.
However, eternal inflation creates infinitely many pocket universes,
each containing a potentially infinite set of observers. The problem
is to define precisely what it means to be {}``typical'' among infinitely
many possible observers and to compute the resulting distribution
of observed parameters~\cite{Linde:1993nz,Linde:1993xx,GarciaBellido:1993wn,Vilenkin:1994ua}.
This has become known as the {}``measure problem'' (see Refs.~\cite{Guth:2000ka,Aguirre:2006ak,Winitzki:2006rn,Guth:2007ng}
for recent reviews and discussions).

Several approaches to the solution of the measure problem have been
proposed. All the existing approaches are based on regulating the
infinite set of possible observers by introducing a cutoff such that
the set of observers becomes finite and large. Once the cutoff is
imposed, distributions of observable parameters are determined by
sampling the regulated set of observers. The distributions of observables
usually converge to a limit when the cutoff is removed. This limit
is taken as the distribution of values observed at a random location
in the multiverse.

The existing measure prescriptions differ in their methods of introducing
the cutoff. One class of measures either completely ignores the growth
of the volume \cite{Starobinsky:1986fx} or counts only the volume
visible to a single randomly chosen world line, as in the causal diamond measure~\cite{Bousso:2006ev,Bousso:2006ge}.
Measures of this class may be called worldline-based measures, in
distinction from volume-based measures that attempt to sample the
entire volume in the spacetime. (A more fine-grained classification
of measure proposals can be found in Refs.~\cite{Aguirre:2006ak,Vanchurin:2006xp}.)
Existing volume-based proposals include the equal-time cutoff~\cite{Linde:1986fd,Linde:1993nz,Linde:1993xx,GarciaBellido:1993wn,Vilenkin:1994ua,GarciaBellido:1994ci,Linde:2006nw,DeSimone:2008bq,Bousso:2008hz,DeSimone:2008if},
the {}``spherical cutoff''~\cite{Vilenkin:1998kr,Vanchurin:1999iv},
the {}``comoving horizon cutoff''~\cite{Garriga:2005av,Easther:2005wi,Vanchurin:2006qp},
the {}``stationary measure''~\cite{Linde:2007nm}, the {}``no-boundary''
measure with volume weighting~\cite{Hawking:2007vf,Hartle:2007gi,Hartle:2008ng},
and the most recent {}``reheating-volume'' measure~\cite{Winitzki:2008yb,Winitzki:2008ph,Winitzki:2008jp} and the boundary measure \cite{Garriga:2008ks}.
The different methods of regularizing the infinite volume of an eternally inflating universe
often give different results. Thus, predictions for observations in the multiverse depend on the choice of the cutoff.

This problem, which was recognized a long time ago \cite{Linde:1993nz,Linde:1993xx,GarciaBellido:1993wn},
still remains unsolved. In the absence of a mathematically unique
cutoff on the set of observers, one evaluates the competing measure
prescriptions on their own merits. A measure prescription is discarded
if it gives pathological results that are clearly in conflict with
observation. Possible pathologies include ``youngness'' paradox~\cite{Guth:2000ka}
and the related incorrect prediction of the CMB temperature~\cite{Tegmark:2004qd},
the {}``Boltzmann brain'' problem~\cite{Dyson:2002pf,Albrecht:2004ke,Page:2005ur,Page:2006dt,Linde:2006nw,Vilenkin:2006qg,Bousso:2006xc,Page:2006ys,Bousso:2007nd,Gott:2008ii},
and incorrect results for local experiments~\cite{Bousso:2007nd}.
It has been possible to rule out some measures on the basis of the presence of these pathologies.

The focus of this paper is to examine the predictions of the {}``stationary
measure''~\cite{Linde:2007nm} in several cases where pathologies
may appear. This measure is an improved version of the class of the
volume-weighted measures proposed in \cite{Linde:1986fd,Linde:1993nz,Linde:1993xx,GarciaBellido:1993wn,Linde:2006nw}.
By explicit calculations we show that the stationary measure does
not suffer from the youngness paradox, does not bias the observed
CMB temperature towards very low or very high values, does not predict
a dominance of ordinary observers by {}``Boltzmann brains,'' and
gives consistent results for local experiments, such as proton decay.

\section{Predictions of the stationary measure}

In a generic scenario of string theory landscape, the stationary measure
prescription ~\cite{Linde:2007nm} can be formulated as follows. All the different possible
pocket universes are labeled by $j=1$, ..., $N$. We consider the
time-dependent volume distribution $V_{j}(t)$ describing the total
volume at time $t$ in pockets of type $j$.

The distribution $V_{j}(t)$ can be found as the solution of the {}``master
equations''~\cite{Garriga:1997ef,Garriga:2005av} \begin{equation}
\frac{d}{dt}V_{j}=\left[3H_{j}^{\beta}-\sum_{i}\Gamma_{j\rightarrow i}\right]V_{j}+\sum_{i}\Gamma_{i\rightarrow j}V_{i},\end{equation}
 Here $\Gamma_{j\rightarrow i}=\lambda_{i\rightarrow j}\frac{4\pi}{3}H_{j}^{\beta-4}$,
and the transition rates $\lambda_{i\rightarrow j}$ per unit time
per unit volume between pockets with the local Hubble rates $H_{j}$
are considered known.

The parameter $\beta$ describes different choices of the time coordinate.
For example, $\beta=1$ corresponds to the proper time, whereas $\beta=0$
corresponds to the time measured in units of the logarithm of the
scale factor $a$ (the {}``$e$-folding time''). 

One then defines the incoming flux $Q_{j}$ \cite{Linde:2006nw} describing
newly created pockets of type $j$ through the equation \begin{equation}
\frac{d}{dt}Q_{j}(t)=\sum_{i}\Gamma_{i\rightarrow j}V_{i}(t).\label{eq:Q equ}\end{equation}
 The total number of ordinary observers created before time $t$ is
proportional to $Q_{j}(t)$ because ordinary observers appear only
near reheating surfaces, which occur only once within each pocket
after that pocket is created. If we are interested in the subset of
observers that have a certain given property, such as a given CMB
temperature, we need to define the corresponding flux $\tilde{Q}_{j}(t)$
and use it instead of $Q_{j}(t)$.

In general, the symbol $j$ may describe not only different vacua
but a particular type of observers, a particular large scale distribution of
time-dependent scalar fields, or a state of geometry generated by cosmological evolution during and after the end of inflation.
Note, for example, that a given CMB temperature might be the same for a number of different states, and one must not combine clearly distinguishable outcomes before the multiverse probabilities are assigned. This point will be better illustrated in the following section when the stationary measure is tested against youngness paradox. For investigation
of these situations, one should use more general types of equations
that include not only the tunneling probabilities but also the description
of slow rolling and quantum diffusion~\cite{Starobinsky:1986fx,Linde:1993nz,Linde:1993xx,LVW}.

The distribution $V_{j}(t)$ asymptotically reaches a stationary regime
\begin{equation}
V_{j}(t)\approx V_{j}^{(0)}\exp\left[3H_{\max}^{\beta}t\right],\label{stationary}\end{equation}
 where $V_{j}^{(0)}$ are constants determined by a particular landscape
model, $H_{\max}$ is of the order of the largest available $H_{j}$ 
\cite{Aryal:1987vn,Linde:1993xx}.
Using Eq. (\ref{stationary}), we integrate Eq.~(\ref{eq:Q equ})
and obtain the asymptotic behavior of $Q_{j}(t)$, 
\begin{equation}
Q_{j}(t)\approx\frac{\exp\left[3H_{\max}^{\beta}t\right]}{3H_{\max}}\sum_{i}\Gamma_{i\rightarrow j}V_{i}^{(0)}.\end{equation}

These results may seem a bit paradoxical. Naively, one could expect
that the volume of different parts on the universe grows at different
rates, $\propto e^{3H_{j}^{\beta}t}$, which depend on the local value of
the Hubble constant (i.e.~on the local value of the cosmological
constant). However, in this class of measures we are interested in
the total volume of the universe and the total number of observers
in all parts of the universe which allow for their existence. The
universe in the eternal inflation scenario is a growing fractal; the
growth of the fractal is mainly due to the growth of the parts of the
universe with the largest available vacuum energy density, which produces
the asymptotic growth rate $\propto e^{3H_{\max}^{\beta}t}$. The parts of
the fractal where the local expansion is slow ($H_{j}\ll H_{\max}$)
will nevertheless grow at the same rate, $\propto e^{3H_{\max}^{\beta}t}$,
due to the tunneling (or slow rolling) of high-$H$ regions towards
the slower-expanding states. For this reason, the total volume of
all parts of the universe grows at the same global rate $e^{3H_{\max}^{\beta}t}$
regardless of the local expansion rate.

The original idea of Refs.~\cite{GarciaBellido:1993wn,Vilenkin:1994ua}
was to relate the probability of living in different states $j$ to
the ratio of the number of observers $N_{j}(t)$, which, in its turn,
may depend either on the ratio of $V_{j}$ or on the ratio of $Q_{j}$.
If we consider a single universe consisting of different exponentially
large parts, as in \cite{Linde:1986fd,Linde:1993nz,Linde:1993xx,GarciaBellido:1993wn}
(rather than a collection of different universes, as in ~\cite{Vilenkin:1994ua}),
then the exponential factors $e^{3H_{\max}^{\beta}t}$ for different $j$
cancel in these ratios.
However, this cancellation occurs only after
the stationary regime is reached. It takes different times $\Delta t_{j}$
until this stationary regime is established for different vacua $j$.
Therefore a comparison of different pockets until the stationarity
is achieved may not be appropriate. Indeed, the difference between the time delays inappropriately rewards the regions $i$ where inflation ends earlier as compared with the regions $j$ where inflation ends later by a huge factor $e^{3H_{\max}^{\beta}\left(\Delta t_{j}-\Delta t_{i}\right)}$ ~\cite{Linde:2007nm}. This factor  is the main reason of the youngness problem, to be discussed in the next section,  and of the exponential sensitivity of the results of the prescriptions of  Refs.~\cite{Linde:1993nz,Linde:1993xx,GarciaBellido:1993wn,Vilenkin:1994ua} to the choice of time parametrization (parameter $\beta$).

To solve these problems, it was proposed in Ref.~\cite{Linde:2007nm}
to {}``reset'' the clock in each of the vacua and to compare $V_{j}(t)$
not at the same time from the beginning of the cosmological evolution,
but at the same time from the beginning of the stationary regime for
each of the vacua. The main idea of this prescription is to replace the  gauge-noninvariant cut-off at a given time $t$ by a cut-off determined by some physical condition, such as the beginning of the regime of stationarity. 

Because of the time resetting, according to the stationary measure
prescription~\cite{Linde:2007nm}, the probability ratio between
pockets $i$ and $j$ is given not by the asymptotic value of the ratio $\frac{N_{i}(t)}{N_{j}(t)}$, but by the ratio
 \bea
\frac{p_{i}}{p_{j}}&=&\lim_{t\rightarrow\infty}\frac{N_{i}(t+\Delta t_{i})}{N_{j}(t+\Delta t_{j})}\nonumber \\ &=&\lim_{t\rightarrow\infty}\frac{N_{i}(t)}{N_{j}(t)}\, e^{3H_{\max}^{\beta}\left(\Delta t_{i}-\Delta t_{j}\right)},\label{eq:stationary def}
\eea
 where $\Delta t_{j}$ are the time delays from the beginning of the
cosmological evolution to the onset of stationarity in the state $j$. The extra factor $e^{3H_{\max}^{\beta}\left(\Delta t_{i}-\Delta t_{j}\right)}$ exactly cancels the problematic $\beta$-dependent factor $e^{3H_{\max}^{\beta}\left(\Delta t_{j}-\Delta t_{i}\right)}$. 

As a result, the probability of any particular outcome of the cosmological evolution is proportional to the probability of each inflationary trajectory (which can be determined e.g. by the probability of quantum tunneling in the corresponding direction) multiplied by the growth of volume during the cosmological evolution ~\cite{Linde:2007nm}.  
This modification can resolve many of the problems associated with
the original probability measures of Refs.~\cite{Linde:1986fd,Linde:1993nz,Linde:1993xx,GarciaBellido:1993wn}
if the time delays $\Delta t_{j}$ are defined appropriately. In the
models where one can ignore inflationary and thermal perturbations,
the time delay $\Delta t_{j}$ is determined only by the duration
of the slow-roll inflation plus any other classical post-inflationary
evolution~\cite{Linde:2007nm}. In this case, one can define $\Delta t_{j}$
as the time where the earliest pocket of type $j$ can be created.

In the next section we will generalize this prescription to the situations where inflationary and thermal perturbations are taken into account. Our goal will be to demonstrate that the properly defined stationary probability measure does not suffer from the youngness problem even with an account taken of perturbations.

\section{No youngness paradox}

The essence of the youngness paradox is the prevalence of pocket universes
created very recently. As an example, one may consider various versions
of the proper time cutoff introduced in \cite{Linde:1986fd,Linde:1993nz,Linde:1993xx,GarciaBellido:1993wn}.
Because of the exponential growth $\propto\exp[3H_{\max}t]$ of the
number of pocket universes created at time $t$, any volume calculation
with a proper time cutoff will yield an exponentially larger number
of pockets that are even slightly younger and have greater energy
density~\cite{Guth:2000ka}. If one evaluates anthropic probabilities
using the proper time cutoff, one may conclude that we should live
at a very large CMB temperature~\cite{Tegmark:2004qd}, in a stark
conflict with observation. (This problem does not appear if one uses
the volume-weighted probability distribution with the scale factor
cutoff \cite{Linde:1993nz,Linde:1993xx,GarciaBellido:1993wn}.)

The stationary measure cures the youngness problem by
explicitly rewarding the {}``older'' pocket universes through the
additional factor $\exp[3H_{\max}\Delta t_{j}]$. Let us first consider homogeneous pocket universes, ignoring inflationary perturbations, and apply this prescription to the task of determining the CMB temperature $T$,
which is an important case where the youngness problem may appear. Then we consider the total
number $N_{j,T}(t_{c})$ of observers that observe the average CMB
temperature $T$ and exist before an asymptotically large cutoff time
$t_{c}$ in pockets of type $j$.

If we do not use a multiverse measure but ask for the probability
per unit proper 3-volume to live in a local pocket of type $j$ with
temperature $T$, we will find a certain probability distribution
$f_{j}(T)$. This distribution is determined by astrophysics and includes
information about the rate of galaxy formation at different times,
as well as about the likelihood of development of life on planets.
We expect that the distribution $f_{j}(T)$ remains unchanged after
applying a multiverse measure, as long as we include only observers
in pockets of fixed type $j$.

The application of the stationary measure requires finding the time
delay until the stationarity is established for the domains of a certain kind. In the original paper \cite{Linde:2007nm}, the stationarity measure was applied to study the abundance of different vacua, where the detailed analysis of cosmological perturbations is not required. However, if we wish to study the distribution of the observables (such as CMB temperature), then different semi-classical realizations of the cosmological evolution, 
which we will call "trajectories", must be considered separately. Each trajectory gives us a unique finial state. We must first find the time delay of stationarity for each such state before the multiverse probabilities can be assigned. After the probability of each trajectory is calculated according to the stationary measure, the trajectories can be grouped together to study the distribution of observables such as CMB temperature. 
 
Once the time delay for all trajectories is found, the prior relative probability to live in the parts of the universe with temperatures $T_1$ and $T_2$, will be given by summing over all trajectories with given properties:
\begin{equation}
\frac{\sum N_{j, T_{1}}(t_{c})}{\sum N_{j,T_{2}}(t_{c})}\sim\frac{f_{j}(T_{1})V(T_{1})}{f_{j}(T_{2})V(T_{2})}.\label{eq:temperatures 1}\end{equation}
Here $V(T_{i})$ stands for the (in some sense average) local growth of the volume of the
parts of the universe until they reach a stationary regime at a temperature
$T_{i}$; note that this quantity does not depend on $t_{c}$ and
$H_{\max}$. Since the evolution to smaller $T$ takes longer time,
in an expanding universe one has $V(T_{1})<V(T_{2})$ for $T_{1}>T_{2}$,
which eliminates the youngness problem. In particular, in a Friedmann
universe the contribution to the probabilities
is given by the factor \begin{equation}
\frac{V(T_{1})}{V(T_{2})}\sim\frac{a^{3}(T_{1})}{a^{3}(T_{2})}\sim\frac{T_{2}^{3}}{T_{1}^{3}}.\label{oldness}\end{equation}

 This factor leads to a very mild preference for smaller temperatures,
i.e.~no youngness paradox. This agrees with the conclusions of Ref.~\cite{Linde:2007nm}
and the results obtained in Ref.~\cite{Bousso:2007nd} for the case
of the calculation of an average $T$ in each pocket universe.

On the other hand, Ref.~\cite{Bousso:2007nd} suggested that the problems may resurface
when one takes into account the fluctuations of the CMB temperature.
The method of the calculation used in  \cite{Bousso:2007nd} was based on the assumption that one should first find all states with a given average temperature $T$ taking into account inflationary perturbations, and then reset the clock simultaneously for all of these states. The average temperature $T$ was calculated using the old, unmodified measure ~\cite{Linde:1993nz,Linde:1993xx,GarciaBellido:1993wn}, which may easily bring back the youngness (or oldness) paradox. 

However,  this procedure contradicts the spirit of the stationary measure  \cite{Linde:2007nm}. According to the prescription described in the previous section, one should consider all distinguishable outcomes of the cosmological evolution, and reset the clock for each of them separately, which removes the spurious exponential dependence on $3H_{\text{max}}t$ and all problems associated with it. Only after that one should take averages and evaluate the most probable  average temperature $T$.

To explain the meaning of this procedure in application to situations where inflationary fluctuations are present, let us consider  slow roll inflation driven by a scalar field $\phi$, accompanied by production of long wavelength quantum fluctuation $\delta \phi$. The wavelength of these perturbations at the moment of their production is ${\cal O}(H^{{-1}})$, and their typical amplitude is ${H\over 2\pi}$.
After $N$ e-folds of inflation they look as a classical scalar field which is homogeneous on scale $H^{-1} e^N$. 

Previously we distinguished between tunneling and slow rolling ending in different vacua, and studied the corresponding time delays. However, a listing of the final vacuum states does not fully describe the final classical configuration of scalar fields and of the resulting geometry. Let us follow a given point $x$ in comoving coordinates, during a slow roll inflation. The field $\phi$ experiences quantum fluctuations of order ${H\over 2\pi}$ during one Hubble time. Let us make a gross simplification and assume that during each Hubble interval $\Delta t=H^{-1}$ the field $\phi$ at  the point $x$ experiences a quantum jump either along the slow-roll motion of the scalar field or against it, with an amplitude ${H\over 2\pi}$. This simplifying assumption is used here only to help explain the main idea of our approach in a simpler way. For simplicity, we will also assume here that inflation is eternal because of metastability of dS vacua in the landscape, but each slow-roll inflationary trajectory has a finite length; a generalization to the eternal slow-roll inflation will be discussed elsewhere ~\cite{LVW}.

If the field jumps down, in the direction of its classical motion,  we will record this fact by writing 1, if it jumps upwards (against the motion) we will write 0. Any sequence of these jumps can be represented in binary code by a sequence on the type 1001011100100...0010, where first digit corresponds to the quantum jump at the beginning of inflation, and the last number describes the quantum jump during the last e-fold of inflation. The $N$-th digit in this sequence describes average properties of space-time geometry and distribution of matter at the end of inflation on scale $H^{-1} e^N$. For example, if the $n$'s digit is 1, then at the end of inflation the average density of matter in a sphere of size $H^{-1} e^n$ around the point $x$ will be smaller that the average density $\rho$ on a larger scale by ${\delta\rho}$, where the density contrast ${\delta\rho\over \rho}$ can be calculated by methods developed in \cite{Mukhanov:1981xt,Hawking:1982cz,Starobinsky:1982ee,Guth:1982ec,Bardeen:1983qw,Mukhanov:1985rz}. Similarly, if the $n$'s digit is 0, then there will be a local increase of density around the point $x$ on scale $H^{-1} e^n$.
These are two distinctly different outcomes for classical post-inflationary geometry, which cannot be erased by perturbations on smaller or greater scales, corresponding to different $n$.

This means that any sequence of the type of 1001011100100...0010 consisting of $N$ digits determines average properties of the space-time geometry near  point $x$ on any of the scales from $H^{-1}$ to $H^{-1} e^N$. If we change $n$'s digit in this sequence, the new sequence will describe a different spacetime geometry on the scale $H^{-1} e^n$. We can use these sequences for describing different realizations of inflationary trajectories taking into account quantum fluctuations. The stationary measure prescription is then used for assigning probabilities to each of the bit sequences.

To increase precision, one may subdivide time into smaller intervals, take into account the fact that the amplitudes of the inflationary fluctuations may have different values,  and calculate the total length (total duration in time) of each realization of an inflationary trajectory at the point $x$. For each of these trajectories, one can calculate the total time delay until inflation ends, or the total number of e-folds of inflation $N$, and the total volume growth $e^{3N}$ during inflation. Each of these trajectories will describe a classical space-time which has {\it different} properties on scales for which the amplitude and the direction of inflationary quantum jumps differ. If we are interested in different outcomes of the cosmological evolution long after inflation, we should calculate the full time delay and the total growth of volume until the final state is reached. 

Therefore our basic rules for resetting time should apply for each of these classes of trajectories, just as they should apply for bubbles of different vacua. Once we realize this, there is no way back; we must reset time separately for each trajectory that brings us to a different state (a state characterized by a different bit sequence). Then reheating occurs at the same (reset) time for all of these trajectories, so smaller or greater temperatures are no longer rewarded or punished by such factors as $e^{3H_{\text{max}}t}$. This means that  neither youngness nor oldness paradox can appear in this approach.

To summarize, the main difference between our approach and that of  \cite{Bousso:2007nd}  is the order in which the multiverse probabilities are assigned and distinct trajectories are combined. From our perspective, one should first define the probabilities for each trajectory according to the stationary measure. Only after the probabilities are assigned one can test the measure against observations by summing the probabilities over all trajectories with common properties such as CMB temperature. 
With our method of calculations, no youngness or oldness paradox appears in the final results because the source of these paradoxes is eliminated for each of the possible outcomes of the cosmological evolution.

\section{No Boltzmann brains}

A formation of a ``Boltzmann brain'' is a spontaneous appearance
of a classical observer (or even a galaxy complete with observers
and the CMB radiation) out of vacuum fluctuations in empty de Sitter
space. The probability of creating a BB of mass $M_{BB}$ in a pocket
of type $j$ during the epoch of $\Lambda$-domination is expected
to be of order 
\begin{equation}
\Gamma_{j}^{BB}\propto\exp\left[-\frac{M_{BB}}{T_{j}}\right]=\exp\left[-2\pi\frac{M_{BB}}{H_{j}}\right],
\end{equation}
 where $T_{j}$ is the de Sitter temperature in pockets of type $j$
and we have neglected sub-exponential factors. An estimate for $M_{BB}\sim100$ kg
in our vacuum with $\Lambda \sim 10^{{-120}}$  yields 
\begin{equation}
\Gamma_{j}^{BB}\propto\exp\left[-10^{70}\right].
\end{equation}
 However, this estimate is valid only if BBs are produced in dS vacuum
with the present value of the cosmological constant. The best chance
for the BBs to be born is in the hot universe, at $T\sim300$K. At
a much greater temperature, they would be born brain-dead, but at
a smaller temperature the probability of their formation is exponentially
smaller. An estimate of the probability of a spontaneous formation
of a BB at $T=300$K gives \begin{equation}
\Gamma_{j}^{BB}(T\sim300\mbox{K})\propto\exp\left[-10^{40}\right].\end{equation}

It is known that the proper time cutoff does not predict any appreciable
number of BBs relative to ordinary observers (OOs)~\cite{Linde:2006nw}.
However, this may be seen as the flip side of the youngness problem
that affects the proper time cutoff. The youngness problem is due
to the fact that the proper time cutoff rewards pockets that reheated
late; these pockets are enormously more numerous than BBs created
at late times in old pockets. The stationary measure compensates the
reward for young pockets by rewarding older pockets in equal measure;
this effectively removes the youngness problem. Let us show that this
compensation does not reward the BBs.

First of all, let us recall the origin of the BB problem. Naively,
the problem arises because ordinary observers (OOs) appear only during
some finite time interval after formation of each pocket, whereas
BBs can appear from vacuum as long as it exists. If the vacuum decay
takes longer than the rate of growth of the universe (i.e.~$\Gamma_{i}<3H_{i}$),
the total number of BBs born in each particular pocket will grow exponentially
with time. Therefore, eventually the total number of BBs will become
much greater than the total number of OOs born in each particular
pocket. This would make ordinary observers highly atypical.

However, this is not the way one should compare OOs and BBs in the
stationary measure. As we already mentioned, the total rate of growth of volume, of the
number of people, of BBs, or of anything else in the universe is \textit{not}
related to the local growth of volume in each particular pocket (which
is proportional to $e^{3H_{i}t}$). Instead, it is determined by the
overall rate of growth of the total volume of the multiverse which
is given by $e^{3H_{\max}t}$.In other words, once the stationarity for the BB production is attained, {\it there is no additional reward accumulating during  the subsequent evolution of the universe}, which is the main source of the Boltzmann brain problem in other probability measures.

Instead of investigation of individual pockets, we must find the time
delays of establishing the stationarity regimes for OOs and BBs, and
compare them, following Eq.~(20) of Ref.~\cite{Linde:2007nm}:
\begin{equation}
\frac{N_{j}^{BB}}{N_{j}^{OO}}=\frac{\Gamma_{j}^{BB}V_{j}(t_{s}^{BB})}{V_{j}(t_{s}^{OO})}\sim\Gamma_{j}^{BB}\left(\frac{T_{s}^{OO}}{T_{s}^{BB}}\right)^{3}.\label{BBOO}
\end{equation}
 Here $t_{s}^{OO}$ is the time corresponding to the moment when the
total number of ordinary observers reaches the stationarity regime,
and $T_{s}^{OO}$ is the temperature of the universe at that time.
Meanwhile $t_{s}^{BB}$ ($T_{s}^{BB}$) is the corresponding time
(temperature) for the BBs. If one ignores quantum and thermal perturbations
then $t_{s}^{OO}$ should be close to the present time, $t_{s}^{OO}\sim O(10^{10})$
years, with $T\sim3$K, whereas the most active epoch of the BB production
corresponds to the time $t_{s}^{BB}$ when $T\sim300$K, as discussed
above. If one takes into account thermal perturbations, then under certain conditions (e.g.~BBs are short-lived) they can be born even earlier, and the stationarity can
be reached at even greater temperature.

In all of these cases, $T_{s}^{OO}\ll T_{s}^{BB}$, and therefore
\begin{equation}
\frac{N_{j}^{BB}}{N_{j}^{OO}}\sim\Gamma_{j}^{BB}\left(\frac{T_{s}^{OO}}{T_{s}^{BB}}\right)^{3}\ll\Gamma_{j}^{BB}\lesssim\exp\left[-10^{40}\right].\end{equation}
 This solves the Boltzmann brain problem for the stationary measure.

In our estimates we dropped out the sub-exponential factors
which are necessary, e.g., for a proper normalization of $\Gamma_{j}^{BB}$
as the transition probability per unit time per unit volume. These
factors are unimportant compared with the enormously small factors
like $\exp\left[-10^{40}\right]$, but it would be satisfying to have
a more detailed derivation of a complete result. We show such a derivation
in Appendix A for a simplified case, ignoring the possibility of BB
formation at high temperature and concentrating on the often discussed
possibility when the BBs are born in pure de Sitter regime. The results
of these calculations confirm our conclusion that there is no BB problem
in the stationary measure.

We should note that in our investigation we considered the simplest assumption that the BBs have mass of the order of 100 kg and are produced in our vacuum with $\Lambda \sim 10^{{-120}}$. One may consider a more speculative version of this problem and study a possibility that a BB can be any computer that can appear in any of the  vacua in string theory landscape and run a program similar to the one that operates in a human brain.  In this case, according to~\cite{DeSimone:2008if}, $\Gamma_{j}^{BB}$ can be much greater,   $\Gamma_{j}^{BB} \sim \exp\, [-10^{A}]$, where $A \sim 20$. This may pose a challenge for the causal diamond measure~\cite{Bousso:2006ev,Bousso:2006ge} and for the  scale factor cutoff measure~\cite{Linde:1993nz,Linde:1993xx,DeSimone:2008bq}, where the solution of the BB problem requires the rate of the BB production to be smaller than the rate of vacuum decay in {\it all } vacua  \cite{Bousso:2008hz,DeSimone:2008if}. This is a very strong requirement, and as of now  we do not know whether this condition can be satisfied. 

Meanwhile, according to~\cite{Linde:2007nm}, the ratio of the BBs in vacuum $j$ and OOs in vacuum $i$,  up to subleading factors,  is given by
\begin{equation}
\frac{N_{j}^{BB}}{N_{i}^{OO}}\sim \frac{\Gamma_{hj}\Gamma_{j}^{BB}}{\Gamma_{hi}  } \lesssim \exp\, [-10^{20}]\ \frac{\Gamma_{hj}}{\Gamma_{hi}  } \ll  {\exp\, [-10^{20}]\over  {\Gamma_{hi}  }},
\end{equation}
where $\Gamma_{hi}$ is the probability of tunneling from the vacua with the maximal energy density to the vacuum $i$.
Therefore the Boltzmann brain problem does not appear in the stationary measure if the probability of tunneling $ \Gamma_{hi}$ to {\it at least  one} of the exponentially large number of anthropically allowed vacua   exceeds $\exp\, [-10^{20}]$. An investigation of the rates of tunneling in the landscape performed in \cite{Linde:2006nw,Ceresole:2006iq,Clifton:2007bn,Freivogel:2008wm} suggests that this condition most probably is met for many of such vacua.

\section{Results of local experiments}

Before discussing the application of the stationary measure to predicting
local experiments, we would like to distinguish three quite different
computational tasks where measure prescriptions are currently being
applied in the literature on multiverse cosmology.

The first task is to compare the abundance of {}``pockets'' (causally
separated domains) containing different physical laws, such as bubbles
of different type of vacuum in the string theory landscape, or superhorizon
domains where physics after reheating is different (this may include
varying particle masses, coupling constants, and the cosmological
constant). One intends to determine the probability of finding oneself
in a region with given physics. We refer to this task as {}``pocket
counting.''

The second task is to predict whether we are natural observers that
resulted from standard cosmology or {}``freak'' observers ({}``Boltzmann
brains'' or BBs) that resulted from thermal fluctuations in a hot
universe or in an empty de Sitter space. This is the {}``BB problem.''

The third task is to predict results of \emph{local experiments} performed
in a given pocket with a given set of local physical laws. Experiments
are distinguished from observations in that experiments can be repeated
under chosen conditions, while observations determine the properties
near our accidental position in the universe that we cannot choose.

The answers to the first and the second questions necessarily involve
a multiverse measure. Answering the third question is possible without
applying the multiverse measure, once a pocket universe is chosen.
For instance, one can (in principle) compute the mean number of galaxies
created at different times after reheating and thus determine the
probability distribution of the observed CMB temperature $f_{j}(T)$
in a randomly chosen galaxy within a pocket of type $j$. Similarly,
one can determine the probability of observing an upward or a downward
spin direction of an electron in a given quantum state, or the probability
of decay of an unstable particle with known half-life. Such predictions
are standard tasks in physics and do not require the consideration
of a multiverse.

Nevertheless, one may try to apply a multiverse measure to the calculation
of probabilities of local experiments and see what happens. As an
example, one may consider an experiment that measures the lifetime
of an unstable particle. If we use the volume distribution in the
proper time cutoff~\cite{Linde:1993xx}, the observers who saw the
particle decay (by chance) a time $\delta t$ earlier will be rewarded
by an additional volume expansion factor $\exp(3H_{\max}\delta t)$,
where $H_{\max}$ is typically close to the Hubble rate of the fastest-expanding
pocket in the landscape, which may be near Planck scale or other highest
energy scale accessible to inflation. The enormous factor $\exp(3H_{\max}\delta t)$
is likely to dominate the probability distribution of outcomes. Hence,
we would have to conclude that every unstable particle decays practically
at once, regardless of the pocket type. This is essentially the youngness
problem in a different guise. The same result is found from the equal-time
cutoff in any other time foliation except the scale factor time
$\tau\equiv\ln a(t)$, where the additional expansion is no longer exponential but is equal merely
to the small Hubble expansion during the experiment, $a^{3}(t_{2})/a^{3}(t_{1})$.
This factor, however, will be very large for the proton decay whose
half-life time is sufficiently large, or for processes in the hot
universe that expands very quickly.

If we use the stationary measure, the bias factor $\exp(3H_{\max}\delta t)$,
which could have lead to problems with the measures with the proper
time cutoff, is identically removed, but the resulting decay rate
still depends on the details of the landscape. A yet another answer
is given by using the {}``comoving'' probability distribution $P_{c}$~\cite{Starobinsky:1986fx,Linde:1986fd}
or the causal diamond measure~\cite{Bousso:2006ev}. According
to both of these latter measures, the most likely observed particle
decay time depends not only on the landscape, but also on the initial
conditions at the beginning of the evolution of the universe.

It may appear that all of these measures violate unitarity and should
be abandoned. However, if one restricts attention to pocket universes
of a single type $j$, it turns out that the stationary measure, as
well as the {}``comoving'' probability distribution and the causal diamond measure, predict correctly the results of local measurements.
It would be interesting to understand whether one may consider a measure
prescription as flawed if it does not yield unmodified probabilities
for local measurements when restricted to pockets of one type. For
a critical discussion of this possibility, see Appendix B.

As a test of the stationary measure in this respect, consider a local
experiment that produces outcomes 1 and 2 with probabilities $p_{1}$
and $p_{2}$. Additionally, we assume that the experiment lasts for
times $\delta t_{1}$ and $\delta t_{2}$ if the two respective outcomes
are obtained. We will now apply the stationary measure to predicting
the results of this experiment.

For simplicity, we restrict ourselves to pockets of one type $j$
and assume that all experiments begin at a fixed time $t_{0}$ after
reheating in each pocket, at a unit rate per 3-volume of reheating
surface. Therefore, the number $N_{1}(t_{c})$ of observers before
time $t_{c}$ who saw the outcome 1 is determined by the total volume
in pockets $j$ of reheating regions where reheating occurred at time
$t_{c}-t_{0}-\delta t_{1}$ or earlier. This volume is equal to $Q_{j}(t_{c}-t_{0}-\delta t_{1})$.
Hence, \begin{equation}
N_{1}(t_{c})=p_{1}Q_{j}(t_{c}-t_{0}-\delta t_{1}).\end{equation}
 Similarly, the number of observers who saw outcome 2 is $N_{2}=p_{2}Q_{j}(t_{c}-t_{0}-\delta t_{2})$.
The stationary measure prescription adjusts these volumes by the factors
$\exp\left[3H_{\max}\delta t_{1}\right]$ and $\exp\left[3H_{\max}\delta t_{2}\right]$;
the time delays are equal to $\delta t_{1}$ and $\delta t_{2}$.
Hence, the result of applying the stationary measure prescription
is \begin{equation}
\frac{N_{1}(t_{c})}{N_{2}(t_{c})}\exp[3H_{\max}(\delta t_{1}-\delta t_{2})]=\frac{p_{1}}{p_{2}}.\end{equation}
 It follows that the stationary measure prescription gives unchanged
results for local experiments, as long as one restricts the calculation
to pockets of fixed type.

\section{Discussion}

Investigation of the probability measure in eternal inflation and
string theory landscape is one of the most challenging problems of modern physics.
It might happen that by identifying a probability measure which does
not lead to paradoxes, we will move towards finding a preferable
basis for a consistent semiclassical description of the multiverse.

One of the intuitively attractive predictions of  the stationary measure is that the
probability to live in a pocket of a given type is proportional to
the growth of volume of the universe during slow-roll inflation \cite{Linde:2007nm}.
This correctly reproduces the standard feature of non-eternal inflation
models in compact universes, where the total number of observers is finite and can
be calculated unambiguously. This result helps to explain why inflation
is desirable even in the situations where its probability is small,
and when it requires fine-tuning of the potential. This measure, unlike
several others, predicts $\Omega=1$, which is confirmed by observations
with ever increasing precision.

This feature of the stationary measure may backfire if there is
an unlimited number of flat directions in the landscape. In that case, one may
argue that most observers should live in the pockets where inflation
is described by models with very flat potentials, which makes inflation
extremely long. In some (but not all) inflationary models, the flattening
of the potential  decreases the amplitude
of perturbations of metric. Some authors call this feature the ``$Q$-catastrophe,''
arguing that it could lead to a prediction of an extremely small amplitude
of perturbations of metric \cite{Feldstein:2005bm,Garriga:2005ee}. We do not know yet whether this  is a real problem; several possible solutions have been already proposed, see e.g. ~\cite{GarciaBellido:1994ci,Linde:2005yw,Hall:2006ff}. For a detailed investigation of this issue one would need to have a much better understanding of inflation in the string theory landscape.

The stationary measure is based on the idea that one should replace the time-dependent cutoff by a  cutoff associated with some invariant  physical properties of the cosmological evolution. It is quite possible that in the future the concept of stationarity will be replaced by something else, which will have a simpler operational meaning. One possible way to generalize our approach is to include a cutoff on the scale of self-reproduction of the universe in the regime of a slow-roll inflation; another possibility is to consider as distinguishable only those outcomes that can be actually verified by direct measurements in a given vacuum. We are planning to consider these and other possibilities in \cite{LVW}. However, our analysis of this measure indicates that already at the present stage of its development it has some important advantages. In this paper we argued that it exhibits neither the youngness paradox, nor the Boltzmann brain problem, nor a bias in predictions for local measurements. In addition, the results obtained using this measure do not depend on initial conditions and show only mild dependence on time parametrization in all of those cases which we were able to study~\cite{Linde:2007nm,LVW}. Therefore we believe that the stationary measure is a viable volume-based multiverse measure which deserves further investigation.

\section*{Acknowledgments}

We dedicate this article to the memory of John Archibald Wheeler,
a great thinker and one of the founders of quantum cosmology. The
authors are grateful to Raphael Bousso, Ben Freivogel, Alan Guth, Mahdiyar Noorbala, 
Misao Sasaki, Navin Sivanandam, Alex Vilenkin, and Alexander Westphal
for helpful discussions. The stay of the authors at the YITP at Kyoto
University was supported by the Yukawa International Program for Quark-Hadron
Sciences. The work by A.L. was supported in part by NSF grant PHY-0756174
and by the Alexander-von-Humboldt Foundation. The work of V.V. was
supported in part by the Transregional Collaborative Research Centre
TRR 33 ``The Dark Universe'' and FQXi mini-grant MGB-07-018. 

\appendix

\section{Abundance of Boltzmann brains\label{AppendixA} }

We will compare the abundance of BBs with that of ordinary observers
according to Eq.~(\ref{eq:stationary def}). For simplicity we confine
attention to the case when BBs are produced during the $\Lambda$-dominated
era, i.e.~we approximate the pocket universe by de Sitter spacetime.
Results will not be qualitatively different if we take into account
the initial stage of the Friedmann expansion.

Let us denote by $V_{j}(t;\mbox{age}>\delta t_{j}^{BB})$ the total
3-volume of pockets of type $j$ that nucleated at time $t-\delta t_{j}^{BB}$
or earlier; we call this the {}``aged'' volume. The nucleation rate
of BBs at time $t$ is then $\Gamma_{j}^{BB}V_{j}(t;\mbox{age}>\delta t_{j}^{BB})$,
where $\Gamma_{j}^{BB}$ is the nucleation rate of BBs per unit 4-volume.
The {}``aged'' volume $V_{j}(t;\mbox{age}>\delta t_{j}^{BB})$ is
produced out of the entire volume $V_{j}(t-\delta t_{j}^{BB})$ at
time $t-\delta t_{j}^{BB}$ through expansion and decays due to tunneling
to other pockets. To keep track of that volume, let us consider the
total volume $V_{j}(t_{0})$ in pockets $j$ at time $t_{0}$, and
let us temporarily disregard tunneling from other pockets into $j$
at $t>t_{0}$ since such tunneling yields {}``younger'' volume.
Hence, at time $t>t_{0}$ some of the initial volume $V_{j}(t_{0})$
will have decayed to other pockets, while the rest of that volume
will have grown due to expansion at the local Hubble rate $H_{j}$.
We temporarily denote the resulting volume by $\tilde{V}_{j}(t;t_{0})$.
In other words, $\tilde{V}_{j}(t;t_{0})$ is the part of the volume
in pockets $j$ at time $t$ that has not tunneled, out of the total
volume present at an earlier time $t_{0}$. We may write the differential
equation for $\tilde{V}_{j}(t;t_{0})$, \begin{equation}
\frac{d}{dt}\tilde{V}_{j}(t;t_{0})=\left[3H_{j}-\sum_{i}\Gamma_{j\rightarrow i}\right]\tilde{V}_{j}(t;t_{0}),\label{eq:VJ equ}\end{equation}
 together with the initial condition\begin{equation}
\tilde{V}_{j}(t_{0};t_{0})=V_{j}(t_{0}).\label{eq:V tilde init}\end{equation}
 Once Eqs.~(\ref{eq:VJ equ})--(\ref{eq:V tilde init}) are solved,
we will have $V_{j}(t;\mbox{age}>\delta t_{j}^{BB})=\tilde{V}_{j}(t;t-\delta t_{j}^{BB})$.
The solution is \begin{equation}
\tilde{V}_{j}(t;t_{0})=V_{j}(t_{0})\exp\left[3\tilde{H}_{j}(t-t_{0})\right],\end{equation}
 where we have defined for brevity \begin{equation}
3\tilde{H}_{j}\equiv3H_{j}-\sum_{i}\Gamma_{j\rightarrow i}.\end{equation}
 Hence we obtain the expression for the {}``aged'' volume, \begin{equation}
V_{j}(t;\mbox{age}>\delta t_{j}^{BB})=V_{j}(t-\delta t_{j}^{BB})\exp\left[3\tilde{H}_{j}\delta t_{j}^{BB}\right].\end{equation}
 The number $N_{j}^{BB}(t)$ of BB observers in pockets of type $j$
at time $t$ is found from the equation \begin{equation}
\frac{d}{dt}N_{j}^{BB}(t)=\Gamma_{j}^{BB}V_{j}(t-\delta t_{j}^{BB})\exp\left[3\tilde{H}_{j}\delta t_{j}^{BB}\right].\end{equation}
 Using the exponential asymptotes of $V_{j}(t)$, we integrate the
above equation to \begin{equation}
N_{j}^{BB}(t)=\frac{\Gamma_{j}^{BB}}{3H_{\max}}V_{j}(t-\delta t_{j}^{BB})\exp\left[3\tilde{H}_{j}\delta t_{j}^{BB}\right].\end{equation}

Finally, we compute the ratio of BBs to normal observers in pockets
$j$ by adjusting the time arguments according to the stationary measure
prescription, \begin{equation}
\frac{N_{j}^{BB}(t+\Delta t_{j}+\delta t_{j}^{BB})}{Q_{j}(t+\Delta t_{j})}=\frac{\Gamma_{j}^{BB}V_{j}^{(0)}}{\sum_{i}\Gamma_{i\rightarrow j}V_{i}^{(0)}}\exp\left[3\tilde{H}_{j}\delta t_{j}^{BB}\right].\end{equation}
 Since the eigenvalue $H_{\max}$ and the coefficients $V_{j}^{(0)}$
satisfy the master equation, \begin{equation}
3H_{\max}V_{j}^{(0)}=\left(3H_{j}-\sum_{i}\Gamma_{j\rightarrow i}\right)V_{j}^{(0)}+\sum_{i}\Gamma_{i\rightarrow j}V_{i}^{(0)},\end{equation}
 we obtain the final probability ratio \begin{equation}
\frac{N_{j}^{BB}(t+\Delta t_{j}+\delta t_{j}^{BB})}{Q_{j}(t+\Delta t_{j})}=\frac{\Gamma_{j}^{BB}}{3H_{\max}-3\tilde{H}_{j}}\exp\left[3\tilde{H}_{j}\delta t_{j}^{BB}\right].\label{eq:ratio ans}\end{equation}
 However, this is not yet the complete answer. The significance of
the probability current $Q_{j}$ is that it gives the amount of new
volume which can be populated by OOs. The number of OOs is proportional
to $Q_{j}$ multiplied by the total number of OOs which are born per
unit volume at the time when the stationarity is reached. Since we
are dealing with enormously large (or small) numbers, for our purposes
it is sufficient to make a rough estimate: $N_{j}^{OO}>Q_{j}H_{j}^{3}$
(more than one observer per horizon). Since $H_{\max}\gg\tilde{H}_{j}$,
it follows that \begin{equation}
\frac{N_{j}^{BB}(t+\Delta t_{j}+\delta t_{j}^{BB})}{N_{j}^{OO}(t+\Delta t_{j})}<\frac{\Gamma_{j}^{BB}}{3H_{\max}H_{j}^{3}}\exp\left[3\tilde{H}_{j}\delta t_{j}^{BB}\right].\label{eq:ratio ans2}\end{equation}
 The last factor exactly reproduces the ratio of the volumes in the second term in Eq.
(\ref{BBOO}). Thus we confirmed our previous calculation of the BB
abundance, up to the sub-exponential factor ${3H_{\max}H_{j}^{3}}$.

\section{Local experiments and the multiverse}

In this appendix we discuss whether the multiverse measure can be
used at all for predicting the results of local experiments.

When computing the probability of being born in a pocket of a given
type, we are essentially talking about the result of the \emph{first}
observation, i.e.~an observation not conditioned to our previous
experience. Using the language of quantum mechanics, we are discussing
a reduction of the wave function of the universe at the moment of
the first observation. Meanwhile, when we consider local experiments,
we are trying to evaluate the probability of a sequence of events,
which requires reducing the wave function twice: once at the beginning
of the experiment, and once at the end. The result of the second experiment
will depend not on the original wave function but on the reduced one,
i.e. on the result of the first experiment. Therefore in general it
may be incorrect to use the original unreduced wave function (or the
original probability distribution) for the prediction of the outcome
of the second experiment.

This may seem to be a scholastic discussion; however, let us give
an example that may show that this issue is relevant. Some time ago
Hawking made a statement that the arrow of time is reversed when a
closed universe reaches its largest size and starts to collapse \cite{Hawking:1985af}.
In other words, when the universe reaches the state of maximal expansion
and begins to contract, the dead will rise from their graves. This
result was highly counter-intuitive, but at the first glance there
could be nothing wrong about it. Indeed, according to the Wheeler-DeWitt
equation~\cite{DeWitt:1967yk}, the wave function of the universe
does not depend on time because the total Hamiltonian of the universe
vanishes. Thus, in the absence of other factors, the wave function
of the universe is a function of the scale factor only. And therefore
the wave function should gradually return to its original state when
the scale factor of the universe turns around and begins to decrease.

However, one needs to be careful in interpreting the wave function
of the universe. If the wave function is independent of time, one
cannot talk about the arrow of time or about the time dependence of
the scale factor. To make sense out of this discussion, one must first
divide the universe into two parts, a macroscopic observer and the
rest of the universe, and then describe the evolution of the universe
from the point of view of the observer~\cite{DeWitt:1967yk}.

Once this division is performed, the wave function of the rest of
the universe depends not only on the scale factor but also on the
time as measured by the observer. Then the simple argument relating
the arrow of time to the scale factor becomes invalid. When an observer
makes experiments describing the rest of the universe, he does not
care that prior to his first observation the wave function was a function
of the scale factor only. Instead, he studies a tremendously reduced
wave function describing a huge universe, where the entropy can only
grow and the arrow of time never turns back. For a closely related
discussion of this issue see e.g.~\cite{Page:1985ei}.

We discussed this example because it may be relevant to the possibility
of describing local experiments using the multiverse measure. Eternal
inflation and jumps between different vacua are possible only because
of quantum effects. We are trying to study this quantum process using
semiclassical tools. If our results appear nonsensical, this may be
an indication that we are using our semiclassical intuition in the
situations where that intuition does not apply.

It may happen that this is the deep reason for the paradoxes which
we sometimes encounter at least with some of the proposed probability
measures. It may also happen that some of these measures can give
us the right answer concerning to probability to live in the universe
where protons may exist. When we find protons around us, we learn
something important about our part of the universe, which strongly
reduces the wave function of the rest of the universe. After that,
instead of using the original non-reduced wave function (which in
our case is supposed to be approximately described by the multiverse
approach), we should use our local laws of physics to describe the
proton decay, just as we should have been using local laws of physics
to study arrow of time in the example described above.


\begin{thebibliography}{99}


\bibitem{Wheeler:1964} J. Wheeler, {\it Relativity, Groups and Topology}, 1963 Les Houches Lectures (Gordon and Breach Science Publishers, New York, 1964).

\bibitem{DeWitt:1967yk}
  B.~S.~DeWitt,
``Quantum Theory of Gravity. 1. The Canonical Theory,''
  Phys.\ Rev.\  {\bf 160}, 1113 (1967).
  
\bibitem{Vilenkin:1982de}
  A.~Vilenkin,
``Creation Of Universes From Nothing,''
  Phys.\ Lett.\  B {\bf 117}, 25 (1982).
  
\bibitem{Hartle:1983ai}
  J.~B.~Hartle and S.~W.~Hawking,
``Wave Function Of The Universe,''
  Phys.\ Rev.\  D {\bf 28}, 2960 (1983).

\bibitem{Linde:1983mx}
  A.~D.~Linde,
``Quantum Creation Of The Inflationary Universe,''
  Lett.\ Nuovo Cim.\  {\bf 39}, 401 (1984).
  
\bibitem{Vilenkin:1984wp}
  A.~Vilenkin,
``Quantum Creation Of Universes,''
  Phys.\ Rev.\  D {\bf 30}, 509 (1984).

\bibitem{Linde:2004nz}
  A.~Linde,
``Creation of a compact topologically nontrivial inflationary universe,''
  JCAP {\bf 0410}, 004 (2004)
  [arXiv:hep-th/0408164].

\bibitem{Linde:1982ur}
  A.~D.~Linde,
``Nonsingular Regenerating Inflationary Universe,''
PRINT-82-0554-CAMBRIDGE, see 
http://www.stanford.edu/$\sim$alinde/1982.pdf (1982).

\bibitem{Linde:1982gg}
  A.~D.~Linde,
``The New Inflationary Universe Scenario,''
in: {\it The Very Early Universe,} Proceedings of 
the Nuffield Symposium, Cambridge 1982, pp. 205-249; 
see http://www.stanford.edu/$\sim$alinde/1983.pdf.

\bibitem{Vilenkin:1983xq}
  A.~Vilenkin,
``The Birth Of Inflationary Universes,''
  Phys.\ Rev.\  D {\bf 27}, 2848 (1983).
  
\bibitem{Linde:1986fd}
  A.~D.~Linde,
``Eternally Existing Self-reproducing Chaotic Inflationary Universe,''
  Phys.\ Lett.\  B {\bf 175}, 395 (1986).
  
\bibitem{Lerche:1986cx}
  W.~Lerche, D.~Lust and A.~N.~Schellekens,
``Chiral Four-Dimensional Heterotic Strings from Selfdual Lattices,''
  Nucl.\ Phys.\  B {\bf 287}, 477 (1987).

\bibitem{Bousso:2000xa}
  R.~Bousso and J.~Polchinski,
``Quantization of four-form fluxes and dynamical neutralization of the
cosmological constant,''
  JHEP {\bf 0006}, 006 (2000)
  [arXiv:hep-th/0004134].
  
\bibitem{Kachru:2003aw}
  S.~Kachru, R.~Kallosh, A.~Linde and S.~P.~Trivedi,
``De Sitter vacua in string theory,''
  Phys.\ Rev.\  D {\bf 68}, 046005 (2003)
  [arXiv:hep-th/0301240].
  
\bibitem{Susskind:2003kw}
  L.~Susskind,
``The anthropic landscape of string theory,''
  arXiv:hep-th/0302219.
  
\bibitem{Douglas:2003um}
  M.~R.~Douglas,
``The statistics of string / M theory vacua,''
  JHEP {\bf 0305}, 046 (2003)
  [arXiv:hep-th/0303194].
  
\bibitem{Guth:2000ka}
  A.~H.~Guth,
``Inflation and eternal inflation,''
  Phys.\ Rept.\  {\bf 333}, 555 (2000)
  [arXiv:astro-ph/0002156].
  
\bibitem{Coleman:1980aw}
  S.~R.~Coleman and F.~De Luccia,
``Gravitational Effects On And Of Vacuum Decay,''
  Phys.\ Rev.\  D {\bf 21}, 3305 (1980).
  
\bibitem{Starobinsky:1986fx}
  A.~A.~Starobinsky, ``Stochastic de Sitter (inflationary) 
stage in the early universe'' (1986), in: {\it Current Topics 
in Field Theory, Quantum Gravity and Strings,} Lecture 
Notes in Physics {\bf 206}, eds. H.J. de Vega and N. Sanchez 
(Springer Verlag 1986) p. 107. 

\bibitem{Aguirre:2006na}
  A.~Aguirre, S.~Gratton and M.~C.~Johnson,
``Measures on transitions for cosmology from eternal inflation,''
  Phys.\ Rev.\ Lett.\  {\bf 98}, 131301 (2007)
  [arXiv:hep-th/0612195].
  
\bibitem{Linde:1993nz}
  A.~D.~Linde and A.~Mezhlumian,
``Stationary universe,''
  Phys.\ Lett.\  B {\bf 307}, 25 (1993)
  [arXiv:gr-qc/9304015].
  
\bibitem{Linde:1993xx}
  A.~D.~Linde, D.~A.~Linde and A.~Mezhlumian,
``From the Big Bang theory to the theory of a stationary universe,''
  Phys.\ Rev.\  D {\bf 49}, 1783 (1994)
  [arXiv:gr-qc/9306035].
  
\bibitem{GarciaBellido:1993wn}
  J.~Garcia-Bellido, A.~D.~Linde and D.~A.~Linde,
``Fluctuations Of The Gravitational Constant In The Inflationary Brans-Dicke
Cosmology,''
  Phys.\ Rev.\  D {\bf 50}, 730 (1994)
  [arXiv:astro-ph/9312039].
  
\bibitem{Vilenkin:1994ua}
  A.~Vilenkin,
``Predictions from quantum cosmology,''
  Phys.\ Rev.\ Lett.\  {\bf 74}, 846 (1995)
  [arXiv:gr-qc/9406010].
  
\bibitem{Aguirre:2006ak}
  A.~Aguirre, S.~Gratton and M.~C.~Johnson,
``Hurdles for recent measures in eternal inflation,''
  Phys.\ Rev.\  D {\bf 75}, 123501 (2007)
  [arXiv:hep-th/0611221].
  
\bibitem{Winitzki:2006rn}
  S.~Winitzki,
``Predictions in eternal inflation,''
  Lect.\ Notes Phys.\  {\bf 738}, 157 (2008)
  [arXiv:gr-qc/0612164].
  
\bibitem{Guth:2007ng}
  A.~H.~Guth,
``Eternal inflation and its implications,''
  J.\ Phys.\ A  {\bf 40}, 6811 (2007)
  [arXiv:hep-th/0702178].
  
\bibitem{Bousso:2006ev}
  R.~Bousso,
``Holographic probabilities in eternal inflation,''
  Phys.\ Rev.\ Lett.\  {\bf 97}, 191302 (2006)
  [arXiv:hep-th/0605263].
  
\bibitem{Bousso:2006ge}
  R.~Bousso, B.~Freivogel and I.~S.~Yang,
``Eternal inflation: The inside story,''
  Phys.\ Rev.\  D {\bf 74}, 103516 (2006)
  [arXiv:hep-th/0606114].
  
\bibitem{Vanchurin:2006xp}
  V.~Vanchurin,
  ``Geodesic measures of the landscape,''
  Phys.\ Rev.\  D {\bf 75}, 023524 (2007)
  [arXiv:hep-th/0612215].
  
\bibitem{GarciaBellido:1994ci}
  J.~Garcia-Bellido and A.~D.~Linde,
  ``Stationarity of inflation and predictions of quantum cosmology,''
  Phys.\ Rev.\  D {\bf 51}, 429 (1995)
  [arXiv:hep-th/9408023].
  
  

  
\bibitem{Linde:2006nw}
  A.~Linde,
``Sinks in the Landscape, Boltzmann Brains, and the Cosmological Constant Problem,''
  JCAP {\bf 0701}, 022 (2007)
  [arXiv:hep-th/0611043].
  
\bibitem{DeSimone:2008bq}
  A.~De Simone, A.~H.~Guth, M.~P.~Salem and A.~Vilenkin,
``Predicting the cosmological constant with the scale-factor cutoff measure,''
  arXiv:0805.2173 [hep-th].
  
    
\bibitem{Bousso:2008hz}
  R.~Bousso, B.~Freivogel and I.~S.~Yang,
``Properties of the scale factor measure,''
  arXiv:0808.3770 [hep-th].
  
\bibitem{DeSimone:2008if}
  A.~De Simone, A.~H.~Guth, A.~Linde, M.~Noorbala, M.~P.~Salem and A.~Vilenkin,
``Boltzmann brains and the scale-factor cutoff measure of the multiverse,''
  arXiv:0808.3778 [hep-th].
  
\bibitem{Vilenkin:1998kr}
  A.~Vilenkin,
``Unambiguous probabilities in an eternally inflating universe,''
  Phys.\ Rev.\ Lett.\  {\bf 81}, 5501 (1998)
  [arXiv:hep-th/9806185].
  
\bibitem{Vanchurin:1999iv}
  V.~Vanchurin, A.~Vilenkin and S.~Winitzki,
``Predictability crisis in inflationary cosmology and its resolution,''
  Phys.\ Rev.\  D {\bf 61}, 083507 (2000)
  [arXiv:gr-qc/9905097].
  
\bibitem{Garriga:2005av}
  J.~Garriga, D.~Schwartz-Perlov, A.~Vilenkin and S.~Winitzki,
``Probabilities in the inflationary multiverse,''
  JCAP {\bf 0601}, 017 (2006)
  [arXiv:hep-th/0509184].
  
\bibitem{Easther:2005wi}
  R.~Easther, E.~A.~Lim and M.~R.~Martin,
``Counting pockets with world lines in eternal inflation,''
  JCAP {\bf 0603}, 016 (2006)
  [arXiv:astro-ph/0511233].
  
\bibitem{Vanchurin:2006qp}
  V.~Vanchurin and A.~Vilenkin,
``Eternal observers and bubble abundances in the landscape,''
  Phys.\ Rev.\  D {\bf 74}, 043520 (2006)
  [arXiv:hep-th/0605015].
  
\bibitem{Linde:2007nm}
  A.~Linde,
  ``Towards a gauge invariant volume-weighted probability measure for   eternal inflation,''
  JCAP {\bf 0706}, 017 (2007)
  [arXiv:0705.1160 [hep-th]].

\bibitem{Hawking:2007vf}
  S.~W.~Hawking,
  ``Volume Weighting in the No Boundary Proposal,''
  arXiv:0710.2029 [hep-th].

\bibitem{Hartle:2007gi}
  J.~B.~Hartle, S.~W.~Hawking and T.~Hertog,
``The No-Boundary Measure of the Universe,''
  Phys.\ Rev.\ Lett.\  {\bf 100}, 201301 (2008)
  [arXiv:0711.4630 [hep-th]].
  
\bibitem{Hartle:2008ng}
  J.~B.~Hartle, S.~W.~Hawking and T.~Hertog,
``The Classical Universes of the No-Boundary Quantum State,''
  Phys.\ Rev.\  D {\bf 77}, 123537 (2008)
  [arXiv:0803.1663 [hep-th]].
  
\bibitem{Winitzki:2008yb}
  S.~Winitzki,
``A volume-weighted measure for eternal inflation,''
  Phys.\ Rev.\  D {\bf 78}, 043501 (2008)
  [arXiv:0803.1300 [gr-qc]].
  
\bibitem{Winitzki:2008ph}
  S.~Winitzki,
``Reheating-volume measure for random-walk inflation,''
  arXiv:0805.3940 [gr-qc].
  
\bibitem{Winitzki:2008jp}
  S.~Winitzki,
``Reheating-volume measure in the landscape,''
  arXiv:0810.1517 [gr-qc].
  
  \bibitem{Garriga:2008ks}
  J.~Garriga and A.~Vilenkin,
``Holographic Multiverse,''
  arXiv:0809.4257 [hep-th].
  
\bibitem{Tegmark:2004qd}
  M.~Tegmark,
``What does inflation really predict?,''
  JCAP {\bf 0504}, 001 (2005)
  [arXiv:astro-ph/0410281].
  
\bibitem{Dyson:2002pf}
  L.~Dyson, M.~Kleban and L.~Susskind,
``Disturbing implications of a cosmological constant,''
  JHEP {\bf 0210}, 011 (2002)
  [arXiv:hep-th/0208013].
  
\bibitem{Albrecht:2004ke}
  A.~Albrecht and L.~Sorbo,
 ``Can the universe afford inflation?,''
  Phys.\ Rev.\  D {\bf 70}, 063528 (2004)
  [arXiv:hep-th/0405270].
  
  
\bibitem{Page:2005ur}
  D.~N.~Page,
``The lifetime of the universe,''
  J.\ Korean Phys.\ Soc.\  {\bf 49}, 711 (2006)
  [arXiv:hep-th/0510003].
  
\bibitem{Page:2006dt}
  D.~N.~Page,
``Is our universe likely to decay within 20 billion years?,''
  arXiv:hep-th/0610079.
  
\bibitem{Vilenkin:2006qg}
  A.~Vilenkin,
``Freak observers and the measure of the multiverse,''
  JHEP {\bf 0701}, 092 (2007)
  [arXiv:hep-th/0611271].
  
\bibitem{Bousso:2006xc}
  R.~Bousso and B.~Freivogel,
``A paradox in the global description of the multiverse,''
  JHEP {\bf 0706}, 018 (2007)
  [arXiv:hep-th/0610132].
  
\bibitem{Page:2006ys}
  D.~N.~Page,
``Return of the Boltzmann brains,''
  arXiv:hep-th/0611158.
  
\bibitem{Bousso:2007nd}
  R.~Bousso, B.~Freivogel and I.~S.~Yang,
``Boltzmann babies in the proper time measure,''
  Phys.\ Rev.\  D {\bf 77}, 103514 (2008)
  [arXiv:0712.3324 [hep-th]].
  
\bibitem{Gott:2008ii}
  J.~R.~I.~Gott,
``Boltzmann Brains--I'd Rather See Than Be One,''
  arXiv:0802.0233 [gr-qc].
  
\bibitem{Garriga:1997ef}
  J.~Garriga and A.~Vilenkin,
``Recycling universe,''
  Phys.\ Rev.\  D {\bf 57}, 2230 (1998)
  [arXiv:astro-ph/9707292].
  
  \bibitem{LVW} A. Linde, V. Vanchurin and S. Winitzki, in preparation.
  
\bibitem{Aryal:1987vn}
  M.~Aryal and A.~Vilenkin,
The fractal dimension of inflationary universe,''
  Phys.\ Lett.\  B {\bf 199}, 351 (1987).
  
\bibitem{Mukhanov:1981xt}
  V.~F.~Mukhanov and G.~V.~Chibisov,
``Quantum Fluctuation And Nonsingular Universe. (In Russian),''
  JETP Lett.\  {\bf 33}, 532 (1981)
  [Pisma Zh.\ Eksp.\ Teor.\ Fiz.\  {\bf 33}, 549 (1981)].
  
\bibitem{Hawking:1982cz}
  S.~W.~Hawking,
``The Development Of Irregularities In A Single Bubble Inflationary Universe,''
  Phys.\ Lett.\  B {\bf 115}, 295 (1982).
  
\bibitem{Starobinsky:1982ee}
  A.~A.~Starobinsky,
``Dynamics Of Phase Transition In The New Inflationary Universe Scenario And
Generation Of Perturbations,''
  Phys.\ Lett.\  B {\bf 117}, 175 (1982).
  
\bibitem{Guth:1982ec}
  A.~H.~Guth and S.~Y.~Pi,
``Fluctuations In The New Inflationary Universe,''
  Phys.\ Rev.\ Lett.\  {\bf 49}, 1110 (1982).
  
\bibitem{Bardeen:1983qw}
  J.~M.~Bardeen, P.~J.~Steinhardt and M.~S.~Turner,
``Spontaneous Creation Of Almost Scale - Free Density Perturbations In An
Inflationary Universe,''
  Phys.\ Rev.\  D {\bf 28}, 679 (1983).
  
\bibitem{Mukhanov:1985rz}
  V.~F.~Mukhanov,
``Gravitational Instability Of The Universe Filled With A Scalar Field,''
  JETP Lett.\  {\bf 41}, 493 (1985)
  [Pisma Zh.\ Eksp.\ Teor.\ Fiz.\  {\bf 41}, 402 (1985)].
  
\bibitem{Ceresole:2006iq}
  A.~Ceresole, G.~Dall'Agata, A.~Giryavets, R.~Kallosh and A.~Linde,
``Domain walls, near-BPS bubbles, and probabilities in the landscape,''
  Phys.\ Rev.\  D {\bf 74}, 086010 (2006)
  [arXiv:hep-th/0605266].
  
\bibitem{Clifton:2007bn}
  T.~Clifton, S.~Shenker and N.~Sivanandam,
``Volume Weighted Measures of Eternal Inflation in the Bousso-Polchinski
Landscape,''
  JHEP {\bf 0709}, 034 (2007)
  [arXiv:0706.3201 [hep-th]].
  
\bibitem{Freivogel:2008wm}
  B.~Freivogel and M.~Lippert,
``Evidence for a bound on the lifetime of de Sitter space,''
  arXiv:0807.1104 [hep-th].

  
\bibitem{Feldstein:2005bm}
  B.~Feldstein, L.~J.~Hall and T.~Watari,
``Density perturbations and the cosmological constant from inflationary landscapes,''
  Phys.\ Rev.\  D {\bf 72}, 123506 (2005)
  [arXiv:hep-th/0506235].
  
  
\bibitem{Garriga:2005ee}
  J.~Garriga and A.~Vilenkin,
``Anthropic prediction for Lambda and the Q catastrophe,''
  Prog.\ Theor.\ Phys.\ Suppl.\  {\bf 163}, 245 (2006)
  [arXiv:hep-th/0508005].
  

\bibitem{Linde:2005yw}
  A.~Linde and V.~Mukhanov,
``The curvaton web,''
  JCAP {\bf 0604}, 009 (2006)
  [arXiv:astro-ph/0511736].
  
\bibitem{Hall:2006ff}
  L.~J.~Hall, T.~Watari and T.~T.~Yanagida,
``Taming the runaway problem of inflationary landscapes,''
  Phys.\ Rev.\  D {\bf 73}, 103502 (2006)
  [arXiv:hep-th/0601028].
  
\bibitem{Hawking:1985af}
  S.~W.~Hawking,
``The Arrow Of Time In Cosmology,''
  Phys.\ Rev.\  D {\bf 32}, 2489 (1985).
  
\bibitem{Page:1985ei}
  D.~N.~Page,
``Will Entropy Decrease If The Universe Recollapses?''
  Phys.\ Rev.\  D {\bf 32}, 2496 (1985).
  
  
\end{thebibliography}
\end{document}